\documentclass[prl,
               twocolumn,
               amsmath,
               amssymb,
               superscriptaddress,
               citeautoscript,
               showpacs,
               floatfix,
]{revtex4}

\usepackage{graphicx}
\usepackage{color}
\usepackage[utf8]{inputenc}
\usepackage[version=3]{mhchem}
\usepackage{xspace}
\usepackage[american]{babel}

\newcommand{\Ti}{\ce{Ti}\xspace}
\newcommand{\Fe}{\ce{Fe}\xspace}
\newcommand{\V}{\ce{V}\xspace}
\newcommand{\Ox}{\ce{O}\xspace}
\newcommand{\La}{\ce{La}\xspace}
\newcommand{\sto}{\ce{SrTiO3}\xspace}
\newcommand{\lvo}{\ce{LaVO3}\xspace}
\newcommand{\lfo}{\ce{LaFeO3}\xspace}

\newcommand{\lao}{\ce{LaAlO3}\xspace}
\newcommand{\lvosto}{\ce{\lvo|\sto}\xspace}
\newcommand{\stolvo}{\ce{\lvo|\sto}\xspace}

\newcommand{\laosto}{\ce{\lao|\sto}\xspace}

\newcommand{\dft}{\textsc{dft}\xspace}
\newcommand{\dftu}{\textsc{dft+u}\xspace}

\newcommand{\wien}{\textsc{Wien}2k\xspace}
\newcommand{\pbesol}{\textsc{pbe}sol\xspace}

\let\oeqref\eqref
\renewcommand{\eqref}[1]{\oeqref{eq:#1}}
\newcommand{\figref}[1]{Fig.~\ref{fig:#1}}

\newcommand{\supref}[1]{Supplemental Fig.~#1}

\newcommand{\optgap}{\ensuremath{\Delta}\xspace}

\newcommand{\eff}{\ensuremath{\eta}\xspace}

\newcommand{\minisec}[1]{\noindent\textbf{\textit{#1}}\hspace{1em}}

\begin{document}

\date{\today}

\author{Elias Assmann}
\affiliation{Institute of Solid State Physics, Vienna University of
  Technology, 1040 Vienna, Austria}
\author{Peter Blaha}
\affiliation{Institute of Materials Chemistry, Vienna University of
  Technology, 1040 Vienna, Austria}
\author{Robert Laskowski}
\affiliation{Institute of Materials Chemistry, Vienna University of
  Technology, 1040 Vienna, Austria}
\author{Karsten Held}
\affiliation{Institute of Solid State Physics, Vienna University of
  Technology, 1040 Vienna, Austria}
\author{Satoshi Okamoto}
\altaffiliation{okapon@ornl.gov}
\affiliation{Materials Science and Technology Division, Oak Ridge
  National Laboratory, Oak Ridge, Tennessee 37831, USA}
\author{Giorgio Sangiovanni}
\affiliation{\mbox{Institut für Theoretische Physik und Astrophysik,
   Universität Würzburg, 97074 Würzburg, Germany}}

\title{Oxide Heterostructures for Efficient Solar Cells}

\begin{abstract}
  We propose an unexplored class of absorbing materials for
  high-efficiency solar cells: heterostructures of transition-metal
  oxides.  In particular, \lvo grown on \sto has a direct band gap
  $\sim{1.1}\:\text{eV}$ in the optimal range as well as an internal
  potential gradient, which can greatly help to separate the
  photogenerated electron-hole pairs.  Furthermore, oxide
  heterostructures afford the flexibility to combine \lvo with other
  materials such as \lfo in order to achieve even higher efficiencies
  with band-gap graded solar cells.  We use density-functional theory
  to demonstrate these features.
\end{abstract}



\maketitle

%


The stunning discovery by Ohtomo and Hwang
\cite{ohtomoNature419,ohtomoNature427} that the interface between the
band insulators \sto and \ce{LaAlO3} (or \ce{LaTiO3}) can become
conductive above a \textit{critical thickness} of the latter opened
the research field of layered oxide heterostructures.  The interest
comes not only from novel physical effects that are absent in the
constituent bulk materials, but also from the perspective  of designing
and tuning specific properties to achieve desired functionalities.
This flexibility becomes even more pronounced when materials with
partially filled $d$-shells, and hence strong electronic correlations,
are involved \cite{okamotoNature428, millis}.  In particular, these
novel oxide heterostructures raise hope to surpass state-of-the-art
semiconductors for specific applications.  In this Letter, we propose
high-efficiency solar cells as such an application.

The performance of photovoltaic systems is characterized by the
efficiency \eff of conversion of the incident photon energy to
electrical power.  A natural limit for \eff is set by the optical gap
\optgap of the absorbing material: photons with energy $\hbar\omega <
\optgap$ are not absorbed at all; those with $\hbar\omega > \optgap$
may be absorbed but each only contributes an amount of energy equal to
\optgap, with the difference being lost to relaxation processes, i.e.,
heat generation.  This consideration, combined with further loss
channels, leads to the famous \emph{Shockley-Queisser limit}
\cite{shockleyJApplPhys510}, which gives an upper bound for \eff as a
function of \optgap with $\eff \alt {34} \%$ in the optimum range
$\optgap = 1\text{--}1.5\:\text{eV}$.  In the case of semiconductors,
a great effort now concentrates on the reduction of so-called
Shockley-Read-Hall electron-hole recombination caused by
defect-induced ``trap'' states and, in general, any intrinsic
recombination mechanism.  Even small improvements on the final
efficiency of a solar cell have an immense economical and
environmental impact, and are intensively pursued by industry
\cite{polmanNatMat11,EPIA}.

\begin{figure}[htbp]
  \includegraphics[width=.76\columnwidth]{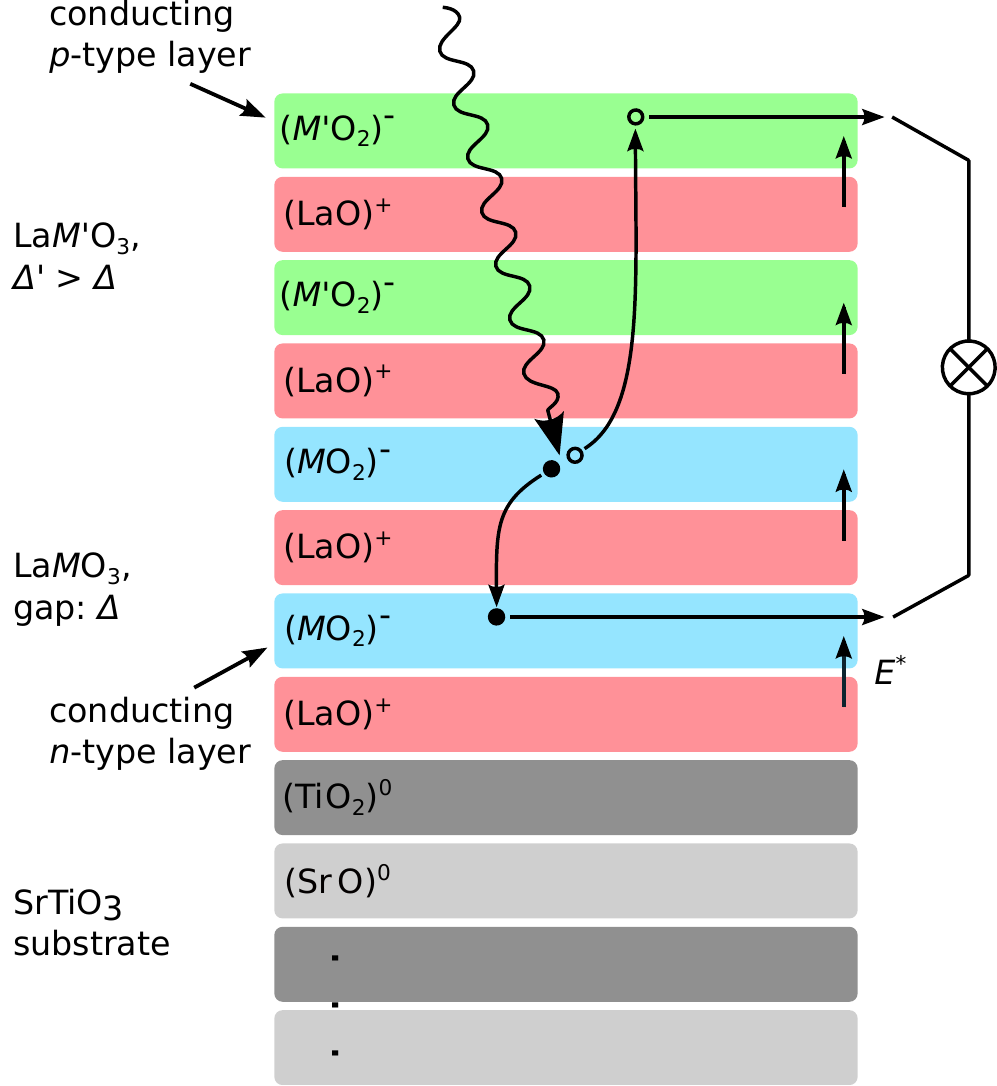}
  \caption{Schematic of a gap-graded oxide-heterostructure solar cell.
    The cell shown here combines two absorber materials, generically
    \ce{La$M'$O3} and \ce{La$M$O3} (\ce{$M$} and \ce{$M'$} being
    different transition metals); the compound with the larger band
    gap $\optgap' > \optgap$ should face the sun.  Incoming photons
    with energy $\hbar\omega > \optgap'$ can be absorbed in the top
    material and contribute an energy $\optgap'$, whereas those with
    $\optgap < \hbar\omega < \optgap'$ traverse the first part, which
    is transparent to them, and can be absorbed in the lower material,
    contributing an energy $\optgap$.  Once an electron-hole pair is
    dissociated, the built-in effective electric field $E^*$ will
    drive the carriers to their respective contacts, which are
    naturally provided in the form of conducting surface and interface
    layers.}
\label{fig:schematic}
\end{figure}

\begin{figure*}[htbp]
  \includegraphics[width=1.1\columnwidth]{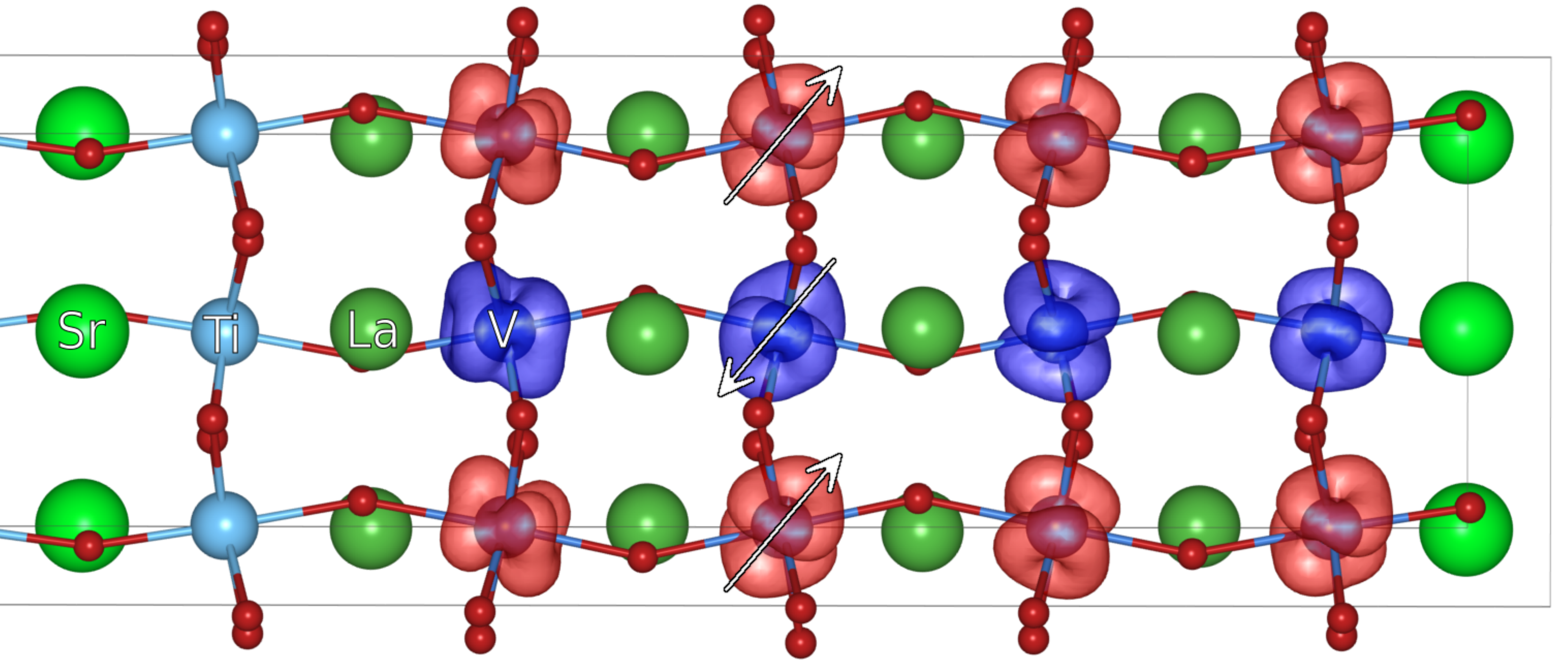}
  \hfill
  \begin{minipage}[b]{.9\columnwidth}
    \caption{Multi-layer \stolvo heterostructure with 4 \ce{VO2}
      planes.  Superimposed on the relaxed structure, the red/blue
      (light/dark gray) lobes show isosurfaces of the
      positive/negative part of the spin-density $n_\uparrow -
      n_\downarrow$.  Owing to the periodic repetition of the cell,
      both an $n$-type (left) and a $p$-type (right) interface appear.
      The substrate is only partly shown; altogether there are six \Ti
      layers.  As in bulk \lvo, the \V-$d$ electrons show AF-C spin
      order; for the central \V layers, the AF-G orbital order is
      similarly preserved.}
  \label{fig:spin_dens}
  \end{minipage}    
\end{figure*}

\minisec{Why oxide heterostructures?}  There are four properties that
make layered oxide heterostructures, and in particular \lvosto, a
promising candidate for efficient solar cells
(cf.~\figref{schematic}): ($i$) an intrinsic electric field emerges in
the photoabsorbing region, which may efficiently separate
photoexcited electrons and holes; ($ii$) the band gap of
$1.1\:\text{eV}$ is direct and in the ideal energy range for
harvesting the sunlight reaching the Earth's surface; ($iii$)
interfaces and surfaces can become metallic on a thickness of about
one unit cell, and those layers naturally allow for extracting the
charge carriers; and ($iv$) one can flexibly combine different
materials \cite{koida2002,bozovic,santamaria,ramesh,lee-christen} and
grow heterostructures with multiple band gaps, e.g., alternate \lvo
with \lfo in order to realize a so-called \emph{band-gap graded}
design \cite{gap-grading}.  By means of density-functional theory
(\dft) calculations we establish these properties of \lvosto and,
given the enormous flexibility offered by oxide heterostructures, we
suggest them as a new candidate for efficient photovoltaic devices.

\laosto can be considered the prototype for the kind of
heterostructure we consider here, and has attracted by far the most
attention \cite{thiel,brinkman,reyren,caviglia,cen}.  However, for our
purposes \lvosto is more suitable since \laosto has a band gap of
several electron volts, much too large for photovoltaics.  Both
heterostructures share a feature which is most interesting in relation
to the electron-hole recombination problem: a polar interface with an
intrinsic electric field.  In \lvo both \La and \V have nominal
valence $+3$ such that the \ce{LaO} planes are positively charged and
the \ce{VO2} planes are negatively charged, inducing an electric field
between these layers (depicted schematically in \figref{schematic};
for quantitative results see \figref{gradient} below).  In the
substrate \sto, on the other hand, \ce{Sr} has nominal valence $+2$
and \Ti has $+4$, such that the \ce{SrO} and \ce{TiO2} planes are both
charge neutral.  Based on these simple considerations one expects a
potential gradient induced by the polar discontinuity between \lvo (or
\lao) and \sto \cite{nakagawaNatMat5}.  \dft calculations for \laosto
show that, in order to avoid this so-called \emph{polarization
  catastrophe}, the polar discontinuity is partially compensated by
electrons transferred from the top of the heterostructure to the
interface, whereas a substantial gradient persists.  Recently, a
finite potential gradient has also been measured experimentally
\cite{singh-ballaNatPhy7}.  That said, such measurements are
challenging, in particular because of the difficulty of controlling
oxygen vacancies in \lao (or, equally, \lvo), which may greatly
influence the gradient.

\begin{figure}[htbp]
  \includegraphics[width=\columnwidth]{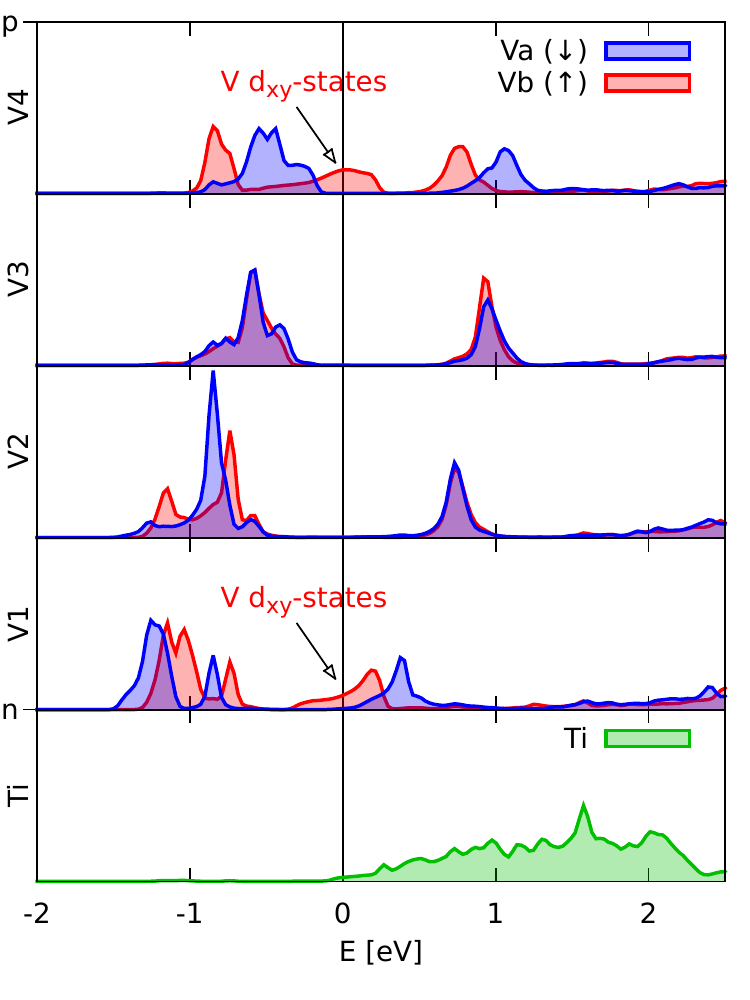}
  \caption{Density of states of a multi-layer \lvosto heterostructure.
    Each of the four \ce{VO2} layers (\V{}1--\V{}4) contains two
    inequivalent \V atoms (\V{}a and \V{}b) carrying opposite spin;
    the majority-spin \V contribution is shown for each atom.  For the
    \Ti contribution (we show only the interfacial layer), the two
    spins as well as the two sites are essentially identical.  These
    are the only relevant contributions around the Fermi level.  Note
    the layer-by-layer shift of the bands which indicates the
    potential gradient, and the appearance of conducting states (of
    $d_{xy}$ character, only in one spin channel) on the interfacial
    \V.}
  \label{fig:DOS_layer}
\end{figure}

\begin{figure*}[htbp]
  \parbox[c]{1.3\columnwidth}{\includegraphics[width=\linewidth]{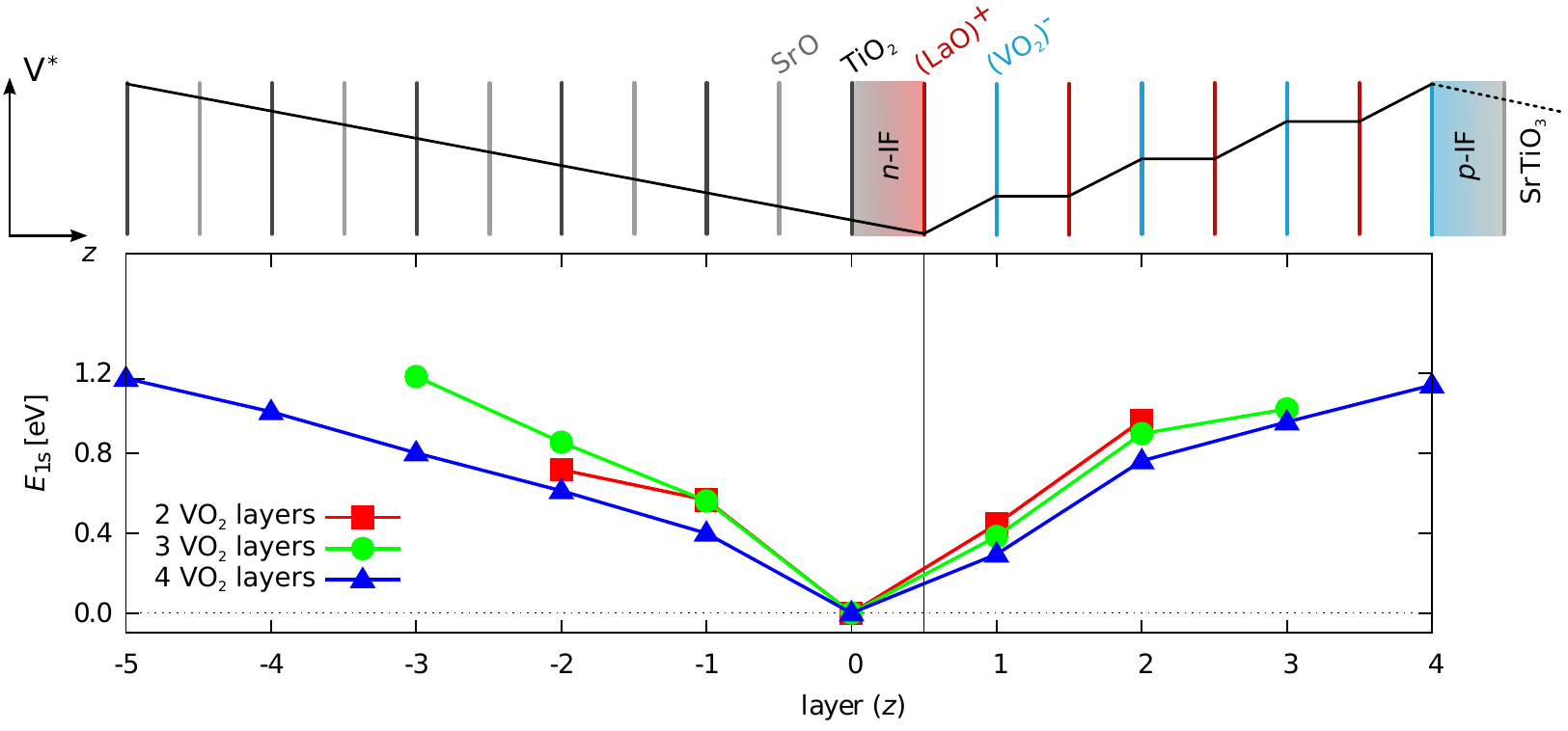}}
  \hfill
  \begin{minipage}[c]{0.7\columnwidth}
    \caption{Potential gradient in multi-layer \lvosto
      heterostructures. The curves
      (\textcolor{red}{\tiny$\blacksquare$}/\textcolor{green}{\large\protect\raisebox{-.15ex}{\textbullet}}/\textcolor{blue}{\footnotesize\protect\raisebox{.12ex}{$\blacktriangle$}}
      distinguish structure size) track the energy of an \Ox-1s state
      through the \ce{TiO2} ($\text{layer} \le 0$) and \ce{VO2}
      ($\text{layer} > 0$) layers, providing a measure of the
      potential gradient.  Above the data, the layers, interfaces
      (IF), and effective electron potential $V^*$ are shown
      schematically.  Periodicity forces the potential to return to
      zero in the \sto part after it has ramped up in the \lvo part.}
    \label{fig:gradient}
  \end{minipage}
\end{figure*}

The built-in electric field will support the separation of
photogenerated electron-hole pairs and will drive the dissociated
charge carriers selectively to the respective contacts for extraction.
\lvosto combines this feature with a near-optimum direct band gap of
$1.1\:\text{eV}$.  Therefore, we concentrate on this structure in the
present paper.

Another advantage of the proposed oxide heterostructures for
solar-cell applications is that electrical contacts to collect the
photogenerated charge carriers are naturally provided.  The
(``$n$-type''; see below) \stolvo interface itself is metallic; in
fact, it would be difficult to contact it otherwise.  Our \dft
calculations show that the surface layer is also metallic and, hence,
suitable to extract the holes.  However, surface defects and edges
might localize the holes.  Hence, careful surface preparation or even
an additional metallic layer such as \ce{SrVO3} or other metallic
contacts will be necessary on the surface side.

\minisec{\textit{Ab initio} simulations of oxide solar cells.}
We study structures with different numbers of \ce{LaO} and \ce{VO2}
layers grown on the (001) surface of \sto considering both ($i$)
periodically repeated arrangements containing one ``$n$-type''
(\ce{TiO2 | [LaO]+}) and one ``$p$-type'' (\ce{SrO | [VO2]-})
interface and ($ii$) an $n$-type (\ce{TiO2 | LaO}) interface with a
slab of vacuum on top of the \lvo part to break periodicity.  These
setups are termed ($i$) ``multilayer'' and ($ii$) ``thin-film''
geometries.

The \dftu calculations were performed using the full-potential
linearized augmented plane-wave code \wien \cite{wien2k}, with the
\pbesol \cite{pbesol} and modified Becke-Johnson \cite{mBJref}
exchange-correlation potentials and a local Coulomb interaction term
$U$ on \Ti, \V and \Fe.  The absorption coefficients in
\figref{absorp} were calculated using the \emph{optic} code for \wien
\cite{wien2kOptic}.  See Supplemental Material \cite{supp} for
technical details, including a discussion of excitonic effects.

We begin by showing one representative \lvosto structure, including
the results from structural optimization and electronic
self-consistency, in \figref{spin_dens}.  As in the bulk
\cite{arima,derayPRL99}, the \V-$d$ electrons show staggered spin and
orbital order.  The spin order is of the C-type (ferromagnetically
coupled antiferromagnetic planes), while the orbital order is of the
G-type (alternating in all directions) \cite{terakura}.  The \V and
\Ti contributions to the density of states for this case are shown in
\figref{DOS_layer}, in the energy range of the \V-$d$ bands.  These
bands shift layer by layer, a first sign of the potential gradient.
The gradient can be evidenced more clearly by tracking the energy of a
core level throughout all layers, as shown in \figref{gradient}.

We estimate the potential slope in the \lvo region to be
$0.08\:\text{eV}/\text{\AA}$ (see \figref{gradient}).  A similar
potential slope is observed in the \lvo region of thin-film structures
(see \supref{1} \cite{supp}).
As the photocarrier excitation takes place in the \lvo region, such a
potential gradient inside \lvo regions helps to efficiently separate
photoexcited electrons and holes.

For the multi-layer heterostructure with four \V layers, we observe
the appearance of states at the Fermi level at the interfaces between
\lvo and \sto.  This confirms the scenario of metallic interfaces due
to the electronic reconstruction found also in \laosto within \dft.
Yet, contrary to \laosto, the carriers reside mostly on the \V rather
than the \Ti, in accordance with recent experimental results
\cite{claessenPrivate}.  The critical thickness that we find for
multi-layer \lvosto heterostructures at the \dftu level is four.  This
result compares well with what has been reported in one experiment
\cite{hottaPRL99} and it is smaller than what was found in another
experiment \cite{claessenPrivate}.  While any solar cell must have a
band gap in order to generate electrical energy, conducting states
that stay confined to the interface may in fact prove useful in
extracting the photogenerated charge carriers.

Let us now turn to a central quantity for solar cell applications, the
optical absorption (i.e. $\alpha$ in the Beer-Lambert law $I(r) \sim
\mathrm{e}^{-\alpha r}$, see Supplemental Material for details
\cite{supp}), shown in \figref{absorp}.  A major advantage of \lvosto
heterostructures over the current standard absorber material is that,
contrary to \ce{Si}, the band gap in \lvo is direct, such that photons
carrying the band-gap energy can create electron-hole pairs without
the aid of phonons or another indirect scattering processes.
Furthermore, across most of the solar spectrum, bulk \lvo compares
favorably with \ce{CdTe}, a direct-gap material currently used for
high-efficiency, thin-film solar cells.

The validity of the absorption coefficient computed within \dft is
confirmed by a comparison to experimental data on bulk \lvo from
Ref. \onlinecite{arima}.  Note that the first sharp peak in the
theoretical curve is very sensitive to details of the crystal
structure (in the limit of a cubic unit cell, it vanishes by
symmetry).  This might explain why it is absent in the experiment and
in the heterostructure calculation.

\begin{figure}[htbp]
  \includegraphics[width=\columnwidth]{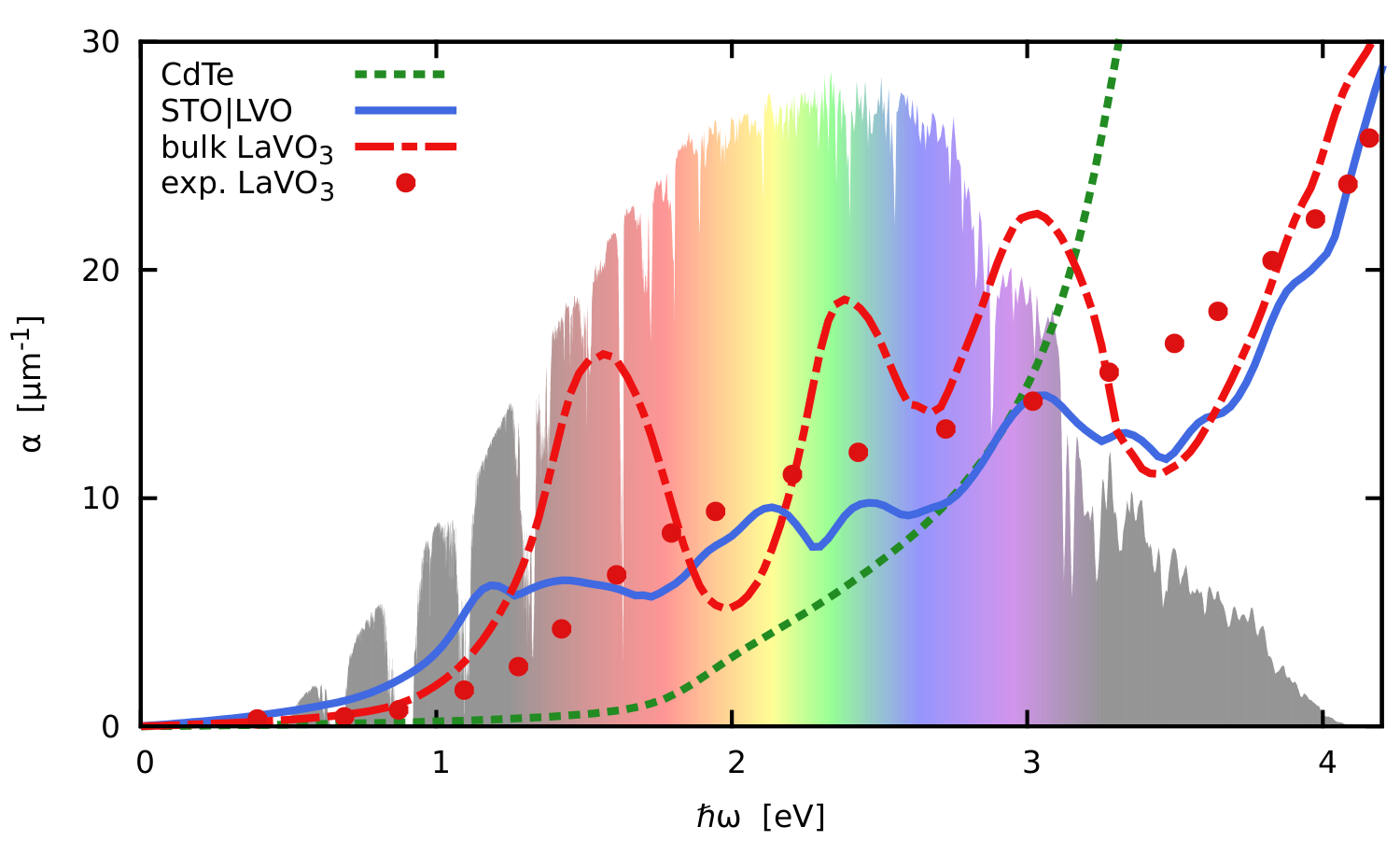}
  \caption{Absorption coefficients and solar spectrum.  The lines show
    the absorption coefficients of bulk \lvo and \lvosto (for a
    multi-layer structure with 2 \V layers) compared to experimental
    data on bulk \lvo \cite{arima} and a calculation for \ce{CdTe},
    which is widely used in current high-efficiency solar cells.  In
    the background, the solar spectrum as measured on the Earth's
    surface is shown (standard global air mass 1.5, in arbitrary
    units).}
  \label{fig:absorp}
\end{figure}

\lvosto heterostructures have similar optical properties as bulk \lvo,
with the additional freedom that individual layers can have different
band gaps owing to layer-dependent distortion and resulting
crystal-field splittings.  Another advantage of oxide heterostructures
is that we can flexibly combine \lvo with a second transition metal
oxide with a somewhat larger band gap, such as \lfo, improving the
conversion efficiency in the high-energy region of the solar spectrum.
Thus one can construct gap-graded structures, in which the layers
facing the sun have larger gaps than the layers underneath.  The \sto
substrate can be on either side because it is transparent to visible
light.  This scheme efficiently reduces the losses due to the
magnitude of the band gap and provides the primary means to exceed the
Shockley-Queisser limit.  We perform \dftu calculations for such a
\ce{\lfo|\lvo|\sto} heterostructure and find a layer-dependent gap of
$2.2\:\text{eV}$ for the \lfo and $1.1\:\text{eV}$ for the \lvo part,
in close agreement with the bulk values.  See~\supref{3} for details
\cite{supp}.

\minisec{Note added in proof.} Manousakis [Phys. Rev. B \textbf{82},
125109 (2010)] reports that, in a Mott insulator, a single photon may
create multiple electron-hole pairs by impact ionization.  This effect
could further increase the efficiency of the \lvosto solar cell
proposed here.

\minisec{Acknowledgments.}  We thank Ralph Claessen, Ho-Nyung Lee,
Margherita Marsili, David Parker, Andrei Pimenov, Jens Pflaum and
Zhicheng Zhong for fruitful discussions, and we acknowledge financial
support from a Vienna University of Technology \emph{innovative
  project} grant (E.A. and G.S.), SFB ViCoM [FWF project ID F4103-N13]
(P. B. and K. H.), and the Laboratory Directed Research and
Development Program of ORNL (S.O.).



\begin{thebibliography}{88}
\bibitem{ohtomoNature419} A. Ohtomo, D. A. Muller, J. L. Grazul and
  H. Y. Hwang, Nature {\bf 419}, 378 (2002).

\bibitem{ohtomoNature427} A. Ohtomo and H. Y. Hwang, Nature {\bf 427},
  423 (2004).

\bibitem{okamotoNature428} S. Okamoto and A. J. Millis, Nature {\bf
    428}, 630 (2004).

\bibitem{millis} A. J. Millis and D. G. Schlom, Phys. Rev. B {\bf 82},
  073101 (2010)

\bibitem{shockleyJApplPhys510} W. Shockley and H. J. Queisser,
  J. Appl. Phys. {\bf 32}, 510 (1961).

\bibitem{polmanNatMat11} A. Polman and H. A. Atwater, Nature
  Mater. {\bf 11}, 174 (2012).

\bibitem{EPIA} EPIA report, ``Global Market Outlook for Photovoltaics
  until 2016''. Downloadable from \url{http://www.epia.org}

\bibitem{koida2002} T. Koida, M. Lippmaa, T. Fukumura, K. Itaka,
  Y. Matsumoto, M. Kawasaki and H. Koinuma, Phys. Rev. B {\bf
    66}, 144418 (2002).

\bibitem{bozovic} I. Bozovic, G. Logvenov, M. A. J. Verhoeven,
  P. Caputo, E. Goldobin and T. H. Geballe, Nature {\bf 422}, 873
  (2003).

\bibitem{santamaria} Z. Sefrioui, D. Arias, V. Pe\~na, J. E. Villegas,
  M. Varela, P. Prieto, C. Leon, J. L. Martinez and J. Santamaria,
  Phys. Rev. B {\bf 67}, 214511 (2003).

\bibitem{ramesh} P. Yu, J.-S. Lee, S. Okamoto, M. D. Rossell,
  M. Huijben, C.-H. Yang, Q. He, J. X. Zhang, S. Y. Yang, M. J. Lee,
  Q. M. Ramasse, R. Erni, Y.-H. Chu, D. A. Arena, Kao, C.-C.,
  L. W. Martin and R. Ramesh, Phys. Rev. Lett. {\bf 105}, 027201
  (2010).

\bibitem{lee-christen} H. N. Lee, H. M. Christen, M. F. Chisholm,
  C. M. Rouleau and D. H. Lowndes, Nature {\bf 433}, 395 (2005).

\bibitem{gap-grading} G. Sassi, J. Appl. Phys. {\bf 54}, 5421 (1983);
M. Konagai and K. Takahashi, J. Appl. Phys. {\bf 46}, 3542 (1975).

\bibitem{thiel} S. Thiel, G. Hammerl, A. Schmehl, Schneider, C. W. and
  J. Mannhart, Science {\bf 313}, 1942 (2006).

\bibitem{brinkman} A. Brinkman, M. Huijben, M. van Zalk, J. Huijben,
  U. Zeitler, J. C. Maan, W. G. van der Wiel, G. Rijnders,
  D. H. A. Blank and H. Hilgenkamp, Nature Mater. {\bf 6}, 493
  (2007).
 
\bibitem{reyren} N. Reyren, S. Thiel, A. D. Caviglia,
  L. F. Kourkoutis, G. Hammerl, C. Richter, C. W. Schneider, T. Kopp,
  A.-S. Ruetschi, D. Jaccard, M. Gabay, D. A. Muller, J.-M. Triscone
  and J. Mannhart, Science {\bf 317}, 1196 (2007)

\bibitem{caviglia} A. D. Caviglia, S. Gariglio, N. Reyren, D. Jaccard,
  T. Schneider, M. Gabay, S. Thiel, G. Hammerl, J. Mannhart and
  J.-M. Triscone, Nature {\bf 456}, 624 (2008).

\bibitem{cen} C. Cen, S. Thiel, G. Hammerl, C. W. Schneider,
  K. E. Andersen, C. S. Hellberg, J. Mannhart and Levy, J.  Nanoscale
  control of an interfacial metal-insulator transition at room
  temperature. Nature Mater. {\bf 7}, 298 (2008).

\bibitem{nakagawaNatMat5} N. Nakagawa, H. Y. Hwang and D. A. Muller,
  Nature Mater. {\bf 5}, 204-209 (2006).

\bibitem{singh-ballaNatPhy7} G. Singh-Bhalla, C. Bell,
  J. Ravichandran, W. Siemons, Y. Hikita, S. Salahuddin, A. F. Hebard,
  H. Y. Hwang and R. Ramesh, Nature Phys. {\bf 7}, 80 (2011).

\bibitem{wien2k} P. Blaha, K. Schwarz, G. K. H. Madsen, D. Kvasnicka,
  and J. Luitz, \emph{WIEN2k, An Augmented Plane Wave + Local Orbitals
    Program for Calculating Crystal Properties} (Techn. Universität
  Wien, Vienna, Austria, 2001). ISBN 3-9501031-1-2

\bibitem{pbesol} J. P. Perdew, A. Ruzsinszky, G. I. Csonka,
  O. A. Vydrov, G. E. Scuseria, L. A. Constantin, X. Zhou and
  K. Burke, Phys. Rev. Lett. {\bf 100}, 136406 (2008)

\bibitem{wien2kOptic} C. Ambrosch-Draxl and J. Sofo,
  Comp. Phys. Comm. {\bf 175}, 1 (2006).

\bibitem{supp} See attached Supplemental Material for technical
  details on the \dft calculations including a discussion of excitonic
  effects, as well as additional results on the gap-graded
  \ce{\lfo|\lvo|\sto} structure and potential gradient in \lvosto.

\bibitem{derayPRL99} M. De Raychaudhury, E. Pavarini and
  O. K. Andersen, Phys. Rev. Lett. {\bf 99}, 126402 (2007).

\bibitem{arima} T.-H. Arima, Y. Tokura, and J. B. Torrance,
  Phys. Rev. B {\bf 48}, 17006 (1993); T.-H. Arima and Y. Tokura,
  Journ. Phys. Soc. Japan {\bf 64}, 2488 (1995).  We calculated the
  absorption coefficient from the reported reflectivity using
  Kramers-Kronig relations following Ref.~\onlinecite{tanner}

\bibitem{terakura} H. Weng and K. Terakura, Phys. Rev. B {\bf 82},
  115105 (2010).

\bibitem{claessenPrivate} R. Claessen (private communication).

\bibitem{hottaPRL99} Y. Hotta, T. Susaki and H. Y. Hwang,
  Phys. Rev. Lett. {\bf 99}, 236805 (2007).

\bibitem{okamotoPRL07} S. Okamoto, A. J. Millis and N. A. Spaldin,
  Phys. Rev. Lett. {\bf 97}, 056802 (2006).

\bibitem{mBJref} F. Tran and P. Blaha, Phys. Rev. Lett. {\bf 102},
  226401 (2009).

\bibitem{tanner} Data analysis package \emph{Datan} by C. Porter and
  D. Tanner, \url{http://www.phys.ufl.edu/~tanner/datan.html}
\end{thebibliography}
\end{document}


\def\figurename{Supplemental Figure}
\def\tablename{Supplemental Table}

\date{\today}

\author{Elias Assmann}
\affiliation{Institute of Solid State Physics, Vienna University of
  Technology, 1040 Vienna, Austria}
\author{Peter Blaha}
\affiliation{Institute of Materials Chemistry, Vienna University of
  Technology, 1040 Vienna, Austria}
\author{Robert Laskowski}
\affiliation{Institute of Materials Chemistry, Vienna University of
  Technology, 1040 Vienna, Austria}
\author{Karsten Held}
\affiliation{Institute of Solid State Physics, Vienna University of
  Technology, 1040 Vienna, Austria}
\author{Satoshi Okamoto}
\affiliation{Materials Science and Technology Division, Oak Ridge
  National Laboratory, Oak Ridge, Tennessee 37831, USA}
\author{Giorgio Sangiovanni}
\affiliation{\mbox{Institut für Theoretische Physik und Astrophysik,
   Universität Würzburg, 97074 Würzburg, Germany}}

\title{Oxide heterostructures for efficient solar cells ---
  Supplementary Information}

\maketitle

\minisec{The DFT calculations} were performed using the full-potential
linearized augmented plane-wave code \wien \cite{wien2k}, with the
\pbesol exchange-correlation potential \cite{pbesol} and a local
Coulomb interaction term $U$ on \Ti, \V and \Fe ($U_\Ti =
9.8\:\text{eV}$, $U_\V = 3\:\text{eV}$, $U_\Fe=5\:\text{eV}$, i.e.~a
\textit{\ggau} scheme.  The $U$ values have been selected to reproduce
the experimental band gaps in the bulk, which is essential in this
case since the heterostructures would become conducting too soon
(i.e.~the critical thickness would be too small) if the gaps were
smaller.

This is also the rationale for the large $U_\Ti$: It is necessary to
shift the \Ti-$d$ states up far enough; but because they are almost
empty, this drastic value is needed to achieve the desired effect.
This is analogous to the case of \ce{LaTiO3}, where a large $U$ on the
empty \ce{La}-$f$ states is necessary to shift them away from
\Ti-$d$.

An independent verification of the results achieved with this $U_\Ti$
comes in the form of a calculation using the parameter-free
\emph{modified Becke-Johnson} (\mbj) exchange-correlation potential
\cite{mBJref}, which we find to qualitatively agree with the
corresponding \dftu calculation.

Formally, periodic boundary conditions are in force in all our
calculations; however, in the thin-film case the vacuum acts to
separate \lvo and \sto, effectively imposing open boundary conditions.

The lattice constants of \sto, \lvo, and \lfo are quite similar when
one takes into account that the unit cells of the latter two are
enlarged with respect to the primitive cubic perovskite cell due to
the reduced symmetry; they are doubled in the $z$-direction, and form
a $\sqrt 2$-cell in the $xy$-plane.  We take this $\sqrt 2$ setting
for the heterostructure unit cell in the $xy$-plane, thus there are
two \Ti/\V/\Fe atoms per plane.  We fix the in-plane lattice parameter
to the value in the \sto substrate (this corresponds to the
experimental situation), while the extent in $z$-direction for the
multi-layer structures corresponds to the sum of the bulk layer
separations.  For the corresponding thin-film structure, this value is
enlarged by $20\,\text{Bohr} \approx 10.6\,\text{\AA}$.  Within this
volume, all atomic positions are relaxed.  For the numerical values,
see \suptab{\ref{supptab:lattc}}.

\begin{table}[htbp]
  \centering
  \caption{Lattice constants of bulk \sto (cubic), \lvo
    (orthorhombic), and \lfo (orthorhombic). $a$ and $b$ are the
    in-plane $\sqrt 2$ lattice parameters, while $c$ is the
    \ce{$B$O2}-plane separation.}
  \label{supptab:lattc}
  \medskip
  \setlength{\doublerulesep}{.4pt}
  \begin{tabular*}{.7\columnwidth}
    {l @{\extracolsep{0pt plus1.5fill}}
      *{3}{r@{\extracolsep{0pt}.}l@{\extracolsep{0pt plus1fill}}}}
    \hline
    &
    \multicolumn{2}{c}{\rule{0pt}{10.5pt}$a/\text\AA$} &
    \multicolumn{2}{c}{$b/\text\AA$} &
    \multicolumn{2}{c}{$c/\text\AA$}
    \\[2pt]
    \hline
    \sto & 5&523 & 5&523 & 3&905 \\
    \lvo & 5&539 & 5&560 & 3&907 \\
    \lfo & 5&554 & 5&568 & 3&928 \\[2pt]
    \hline
  \end{tabular*}
\end{table}

In the bulk, both \lvo and \lfo show antiferromagnetic (AF) spin and
orbital order at low temperatures: \lvo has AF-C spin and AF-G orbital
order while \lfo has AF-G order in both channels; \dft predicts this
in accordance with experiment.  This raises the question which spin
order prevails in the heterostructures.  For \lvosto, we focus on AF-C
both in analogy to the bulk, and because it has the lowest energy in
\dft (among AF-A, AF-C, AF-G and FM) in two test cases.

When a second magnetic material comes into play with \lfo, the
situation becomes more complicated and three different spin orders
have to be considered in general: within \lvo, within \lfo and at the
interface.  In the \ce{\lfo|\lvo|\sto} calculation presented here,
there is only one layer of each material, therefore we need to consider
only the coupling between \lfo and \lvo, which we choose
antiferromagnetic in the interest of comparability with the \lvo-only
calculations, resulting in ``AF-C'' spin order across \V and \Fe.

The absorption coefficient for \lvo was calculated within \ggau as
described above, using the program of Ref. \onlinecite{wien2kOptic}
while for the semiconductors the \mbj potential was used
\cite{mBJref}, which yields band gaps in good agreement with the
experimental values in these cases.  The calculated absorption
coefficients have been checked regarding convergence with respect to
the density of $k$-points used to sample the Brillouin zone.  For the
heterostructure, up to $10 \times 10 \times 1$ $k$-points were used.

The electron-hole interaction, which we investigated by solving the
Bethe-Salpeter equation for bulk \lvo, introduces a sizable ($\sim
0.6\,\text{eV}$) but rigid shift of the spectrum to lower energies.
However, since the shape of the optical absorption is hardly changed,
a somewhat larger value of $U_\V$ could easily compensate this shift,
reproducing once again the experimental value for the optical gap as
in Fig.~5 in the main paper.  Thus we conclude that electron-hole
interactions do not change our predictions significantly.

\begin{figure*}[htbp]
  \parbox[c]{1.3\columnwidth}{\includegraphics[width=\linewidth]{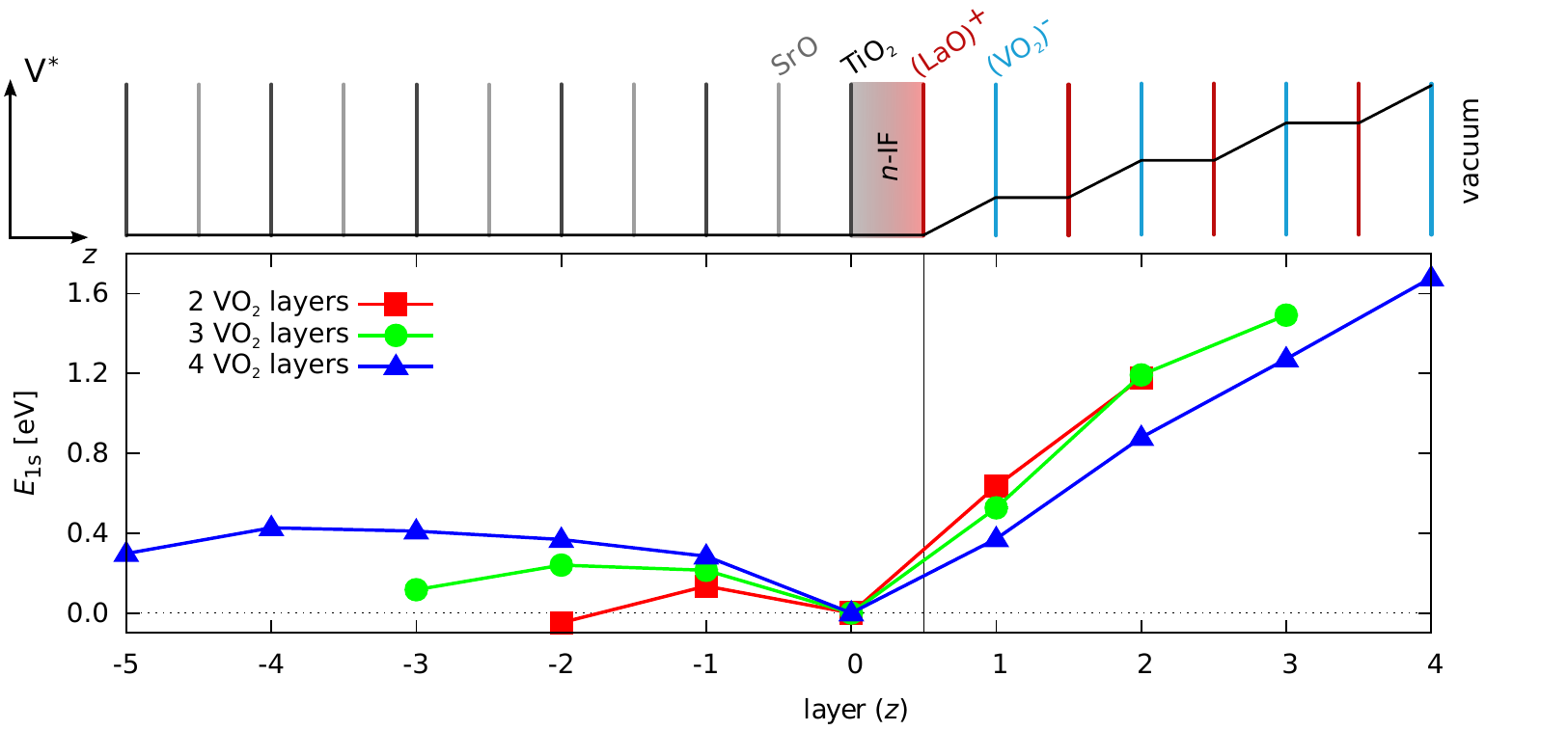}}
  \hfill
  \begin{minipage}[c]{0.7\columnwidth}
    \caption{Potential gradient in thin-film \lvosto
      heterostructures. The curves
      (\textcolor{red}{\tiny$\blacksquare$}/%
      \textcolor{green}{\large\protect\raisebox{-.15ex}{\textbullet}}/%
      \textcolor{blue}{\footnotesize\protect\raisebox{.12ex}{$\blacktriangle$}}
      distinguish structure size) track the energy of an \Ox-1s state
      through the \ce{TiO2} ($\text{layer} \le 0$) and \ce{VO2}
      ($\text{layer} > 0$) layers, which provides a measure of the
      potential gradient.  Above the data, the layers, interfaces (IF),
      and effective electron potential $V^*$ are shown schematically.
      The potential in the \sto part is almost flat, showing only a
      slight buckling which we attribute to band bending at the
      interface.}
    \label{supfig:gradient}
  \end{minipage}
\end{figure*}

The experimental absorption coefficient for \lvo was calculated from
the reflectivity measured in Ref. \onlinecite{arima} using
Kramers-Kronig relations, employing the 
{\em datan} program package \cite{tanner}.

\minisec{Gradient in the thin-film case.}  In analogy to Fig.~4 in the
main paper, we show the behavior of the \Ox-1s level in thin-film
heterostructures in \supref{\ref{supfig:gradient}}.

\begin{figure}[htbp]
  \centering
  \includegraphics[width=\columnwidth]{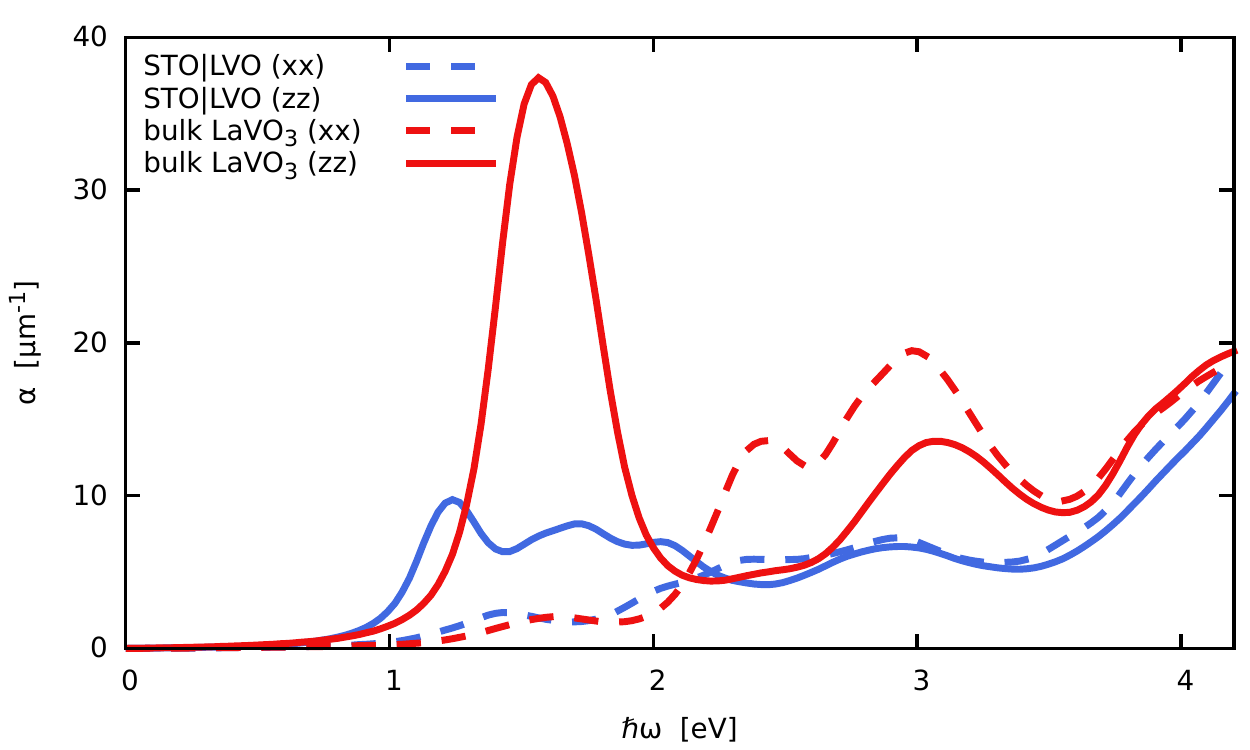}
  \caption{Contributions to absorption coefficients.  Both in bulk
    \lvo and the heterostructure, a significant asymmetry is seen
    between the $xx$ and the $zz$ components, while $yy$ is similar to
    $xx$.  The off-diagonal components are rather small in comparison
    with the diagonal ones.}
  \label{supfig:absorp_detail}
\end{figure}

\minisec{Absorption coefficients.}  As stated in the main text, the
absorption coefficient $\alpha(\omega)$ is defined via the exponential
dampening of the light intensity as a function of distance travelled
through the material.  However, in addition to the frequency
dependency, the absorption also depends on the light's polarization,
leading to a matrix $\alpha_{ij}$ analogous to the dielectric tensor.
In cubic crystals (such as \ce{CdTe} and \ce{GaAs}), symmetry reduces
the tensor to only one independent component, $\alpha_{ii}$, and this
is what is plotted for these materials in Fig.~5 in the main paper.

On the contrary, in \lvo and \lvosto, the reduced symmetry allows six
independent entries.  Which of these may contribute to the actual
absorption depends of course on the light's angle of incidence with
respect to the crystal.  Since the heterostructure is grown along the
$z$-direction, in the first approximation the polarization will be in
the $xy$ plane.  However, because some of the light will come from
other directions due to scattering, and because heterostructures are
often grown on a stepped substrate, some $z$-component will also be
present, and all $\alpha_{ij}$ may contribute.  Therefore, we plotted
a suitably averaged quantity $\bar\alpha := \frac13(\sum_i \alpha_{ii}
+ \sum_{ij} \alpha_{ij})$ in Fig.~5.
\supref{\ref{supfig:absorp_detail}} shows some of the individual
contributions.  The optical absorption for the heterostructure was
calculated in the multi-layer geometry with two \ce{VO2} layers.

\begin{figure}[htbp]
  \includegraphics[width=\columnwidth]{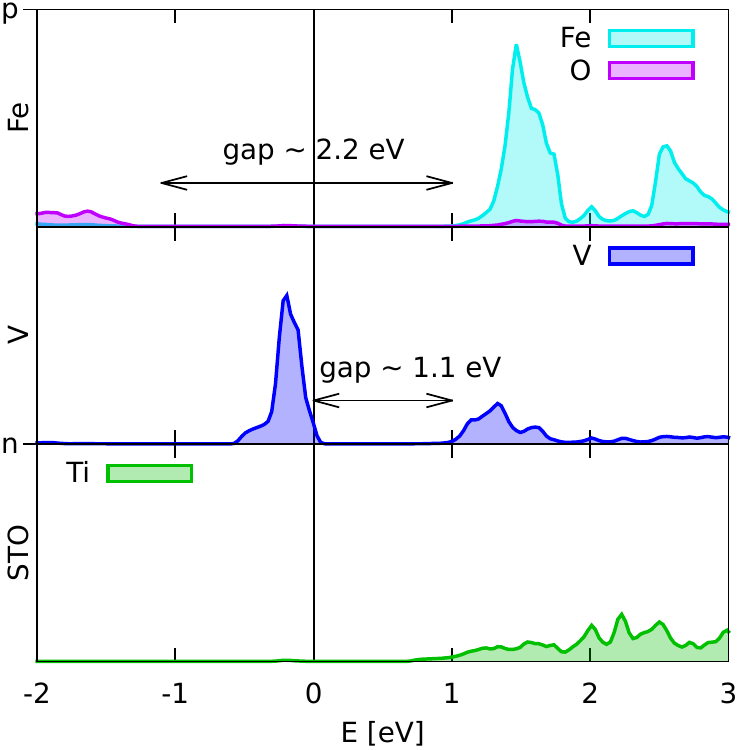}
  \caption[\ce{\lfo|\lvo|\sto} DOS]{
    The relevant contributions to the density of states around the
    Fermi level for the ``gap-graded'' \ce{\lfo|\lvo|\sto}
    structure.  In this case, the two \V/\Fe sites give almost
    identical contributions (in opposite spin channels).  A $p$- and
    an $n$-type interface appear, as marked, due to periodic boundary
    conditions.
  }
  \label{supfig:DOS_layer_Fe}
\end{figure}

\minisec{Band-gap grading with \lfo.} In this section, we give details
on the calculation combining \lvo and \lfo referenced in the main
text.  The structure contains 3 \Ti, 1 \V and 1 \Fe layer, with two
metal atoms per layer.  The projected density of states for \Fe, \V
and the \Ti layer at the $n$-type interface is shown in
\supref{\ref{supfig:DOS_layer_Fe}}.  The differing magnitudes of the
\Fe--\Ox and \V--\V gaps are clearly seen.

It should be emphasized that $\lvo+\lfo$ is only one example for many
different combinations that may be envisaged within the flexible
framework provided by oxide heterostructures.

\minisec{Acknowledgments.}  We thank Andrei Pimenov for the calculation
of the experimental LaVO$_3$ absorption.